\RequirePackage{ifpdf}
\documentclass{PoS}
\usepackage{colordvi}
\usepackage{multirow}
\usepackage{amsfonts,epsfig,cite,graphics}
\usepackage{amssymb}
\usepackage{amsmath}
\usepackage{mathrsfs}
\usepackage{amsbsy}
\usepackage{latexsym}
\usepackage{graphicx}
\usepackage{psfrag}
\usepackage{macros_static}
\usepackage{slashed}
\usepackage[multidot]{grffile}

\newcommand{\arxiv}[1]{arXiv:\,\href{http://arxiv.org/abs/#1}{{\tt #1}}}

\title{First results for SU(2) Yang-Mills with one adjoint Dirac Fermion}
\ShortTitle{First results for SU(2) Yang-Mills with one adjoint Dirac Fermion}

\author{\speaker{Andreas Athenodorou}, \\
Department of Physics, Swansea University, Singleton Park, Swansea SA2 8PP, UK \\
\& Department of Physics, University of Cyprus, Nicosia, Cyprus \\
E-mail: \email{athenodorou.andreas@ucy.ac.cy}}

\author{Ed Bennett, \\
Department of Physics, Swansea University, Singleton Park, Swansea SA2 8PP, UK \\
E-mail: \email{pyedward@swan.ac.uk}}

\author{Georg Bergner, \\
Universit\"at Frankfurt, Institut f\"ur Theoretische Physik Max-von-Laue-Str. 1, D-60438 Frankfurt am Main, Germany \\
E-mail: \email{g.bergner@uni-muenster.de}}

\author{Biagio Lucini, \\
Department of Physics, Swansea University, Singleton Park, Swansea SA2 8PP, UK \\
E-mail: \email{B.Lucini@swan.ac.uk}}

\author{Agostino Patella \\
School of Computing and Mathematics, Plymouth, PL4 8AA, UK \\
\& PH-TH, CERN, CH-1211 Geneva 23, Switzerland, \\
E-mail: \email{agostino.patella@plymouth.ac.uk} }

\abstract{
We present a first exploratory study of SU(2) gauge theory with one Dirac flavour in the
adjoint representation. We provide initial results for the spectroscopy and the anomalous dimension 
for the chiral condensate. Our investigation indicates that the theory is conformal or near-conformal,
with an anomalous dimension of order one. A discussion of the relevance of these findings in relation to walking technicolor scenarios is also presented.}

\FullConference{The XXXI International Symposium on Lattice Field Theory\\
         July 29 - August 03 2013\\
         Mainz, Germany}

\begin{document}
\vspace{-0.30cm}
\section{Introduction}
\label{sec:introduction}
\vspace{-0.35cm}
Novel strong interactions might provide an explanation for the mechanism of electroweak symmetry breaking. In order for new strong dynamics to be able to explain this breaking, the theory must be near the onset of the conformal window and possess an anomalous dimension of the chiral condensate of order one. Numerical lattice simulations provide a crucial non-perturbative tool to check possible candidate models realising these ideas (see~\cite{kuti} for a recent review). Recent lattice calculations~\cite{su2nf2_1,hietanen_1,su2nf2_2,su2nf2_3,degrand,su2nf2_4,giedt} have demonstrated that the SU(2) gauge theory with two adjoint Dirac flavours ($N_f=2$) is in the conformal window. Besides the conformal strong dynamics inducing the electroweak symmetry breaking, an important second requirement for a realistic scenario is an anomalous dimension close to one. The most recent measurements for this theory have revealed an anomalous dimension $\gamma_* \sim 0.37(2)$~\cite{Patella,deldebbio}, thus, ruling out the possible phenomenological relevance of SU(2) with two adjoint flavours.

Hence, it is important to understand whether large anomalous dimensions can arise in the context of conformal gauge theories. Although the anomalous dimension is small at the perturbative zeros of the beta function, large anomalous dimensions might arise near or at the lower end of the conformal window. Since, according to numerical evidence, in the two-flavour case the SU(2) gauge theory is infrared conformal with a small anomalous dimension, the only remaining possibility for observing a large anomalous dimension in SU(2) with adjoint Dirac flavours is to move to the single flavour case. According to some heuristic arguments this theory is considered to be confining; nevertheless, a non-perturbative study from first principles has never been performed. Besides, at large $N$, studies of the infrared behaviour of the theory based on reduction are inconclusive~\cite{bringoltz,narayanan}. 

In these proceedings we present an investigation of the SU(2) Yang-Mills theory with one adjoint Dirac fermion using numerical Monte Carlo studies of the theory discretised on a spacetime lattice. The chiral symmetry breaking of this model produces only two Goldstone bosons, making it insufficient to give mass
to the $W^{\pm}$ and $Z$ bosons. Thus, even if the theory turns out to be in the near-conformal regime, it  has no phenomenological importance. However, it may still provide useful information about the possibility of having a large anomalous dimension.
\vspace{-0.30cm} 
\section{Lattice Calculation}
\vspace{-0.30cm}
\subsection{The Setup}
\label{sec:setup}
%\vspace{-0.30cm}
The lattice formulation used in this work for the action $S$ of the model is the Wilson gauge action $S_G$ and Wilson fermionic action $S_F$:  $S=S_{G} + S_{F}$ with
\begin{eqnarray}
  S_G = \beta \sum_{p} {\rm Tr} \left[ 1 - U(p)  \right] \ \ \ \ {\rm and} \ \ \ \    S_{F} =  \sum_{x,y} {\overline \psi}(x) D(x,y) \psi(y),
\end{eqnarray}
where $D(x,y)$ the massive Dirac operator:
\begin{eqnarray}
  D(x,y) = \delta_{x,y} - \frac{\kappa}{2} \left[  \left(1-\gamma_{\mu}  \right) U_{\mu} (x) \delta_{y,x+\mu}  +  \left(1+\gamma_{\mu}  \right) U^{\dagger}_{\mu} (x-\mu) \delta_{y,x-\mu}   \right] \ .
\end{eqnarray}
$\kappa$ is referred to as the hopping parameter and is related to the bare fermion mass $m$ through $\kappa = 1/(8+2am)$. $\beta = 2 N / {g^2} = 4 / {g^2}$ is the inverse coupling and $N$ the number of colors. 

We have applied the Rational Hybrid Monte Carlo (RHMC) algorithm to produce the gauge configurations. Our code is based on the HiRep suite~\cite{HiRep}. More details about the production of configurations and their analysis will be provided in~\cite{forthcoming}. 
\vspace{-0.30cm}
\subsection{The Phase Diagram}
\label{sec:phasediagram}
\begin{figure}
  \begin{center}
	\includegraphics[width=8cm]{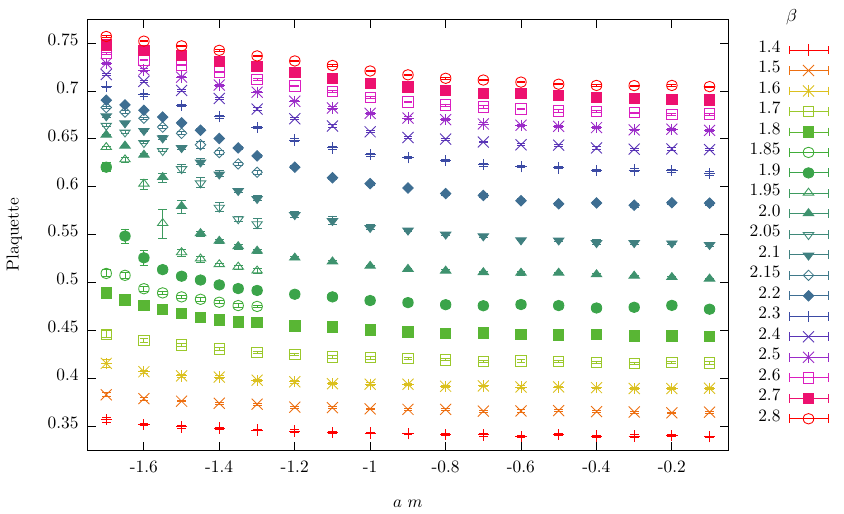}
  \end{center}
        \vspace{-0.6cm} 
	\caption{The results of the pilot study, showing the average plaquette across a range
of value of $\beta$ and $am$; the region of interest was identified around $\beta = 2.05$, $am = -1.5$.
\label{fig:phasediagram}}
%\vspace{-0.45cm} 
\end{figure}

\vspace{-0.20cm}
Since this system had not been investigated before, we performed a careful scan in the bare parameters. Based on related theories, we made a first guess as to where the region of interest would be. The next step was to narrow down the parameter ranges to make the simulation of interesting physics feasible. The lattice phase diagram was explored by investigating the average plaquette on a $4^4$ lattice, choosing the ranges $1.4 \le \beta \le 2.8$, $ - 1.7 \le am \le  - 0.1$, in steps of $am=0.1$. This is presented in Figure~\ref{fig:phasediagram}. Once the region of the bulk phase transition was identified to be around $\beta \simeq 1.9$, $a m \simeq -1.65$, points were added in its neighbourhood to increase the resolution to $0.05$.

Based on these results, a single value of the lattice spacing was chosen corresponding to $\beta = 2.05$, and the bare fermion mass in $ - 1.523 \le am \le  - 1.475$, with $am =  - 1.523$ being our closest point to the chiral limit. For the quantitative measurements that follow, we have chosen lattices of size $N_T \times N^3=$$16 \times 8^3$, $24 \times 12^3$,  $32 \times 16^3$ and $48 \times 24^3$. The parameters for our ensembles are presented in Table~\ref{tab:1}. At each bare mass, the size of the lattice has been fixed by requiring that spectral observables are not affected by finite size artefacts.
\begin{table}[htbp]
\small
\centering
\begin{tabular}{c | c | c c | c ||| c | c | c c | c }
\hline
Lattice	&	$V$	        &	$-am$	&   $a m_{\rm PCAC}$      &  $N_\mathrm{conf}$ & Lattice	&	$V$	        &	$-am$	&   $a m_{\rm PCAC}$      &  $N_\mathrm{conf}$	\\
\hline
A1      &	$16\times8^3$	&	$1.475$	&	0.1489(9)       &	2400  & C2	&	$32\times{16}^3$	&	$1.490$	&	0.1279(2)&       2300	\\	
A2	&	$16\times8^3$	&	$1.500$	&	0.1101(12)      &	2200  & C3	&	$32\times{16}^3$	&	$1.510$	&	0.09111(31)&       2200	\\
A3	&	$16\times8^3$	&	$1.510$	&	0.0904(14)	&	2400  & C4	&	$32\times{16}^3$	&	$1.510$	&	0.09048(52)&       2300	\\
A4	&	$16\times8^3$	&	$1.510$	&	0.0872(22)	&	4000  & C5	&       $32\times{16}^3$	&	$1.514$	&	0.08223(34)&       2300	\\
\cline{1-5}
B1	&	$24\times{12}^3$	&	$1.475$	&	0.1493(5)&	2400  & C6	&	$32\times{16}^3$	&	$1.519$	&	0.06587(37)&       2300	\\
B2	&	$24\times{12}^3$	&	$1.500$	&	0.1113(8)&	2300  & C7	&	$32\times{16}^3$	&	$1.523$	&	0.04840(54)&       2200	\\
\cline{6-10}
B3	&	$24\times{12}^3$	&	$1.510$	&	0.09226(92)&	4000  & D1	&	$48\times{24}^3$	&	$1.510$	&	0.09130(27)&       1534	\\
\cline{1-5}
C1	&	$32\times{16}^3$	&	$1.475$	&	0.1485(4)&       2100 & D2	&	$48\times{24}^3$	&	$1.523$	&	0.04722(43)&       2168	\\
\hline
\end{tabular}
\caption{The lattices considered in this study with their volumes, bare masses, PCAC masses and number of configurations.}
\label{tab:1}
\end{table}

\vspace{-0.45cm} 
\subsection{Observables}
\label{sec:observables}
Our goal in this project is to investigate the infrared regime of the theory. In particular, we are interested in conformal properties and the chiral condensate anomalous dimension. To this purpose, following~\cite{su2nf2_2,su2nf2_3}, we have extracted the mass spectrum in the chiral regime and tested finite size scaling predictions. In addition, we have determined the Dirac mode number as a function of the Dirac eigenvalues~\cite{Patella}. 

The massless fermionic action has an SU(2) global chiral symmetry. In the presence of a nonzero condensate this breaks to $\textnormal{SO(2)}\cong \textnormal{U(1)}$. The generator of the unbroken group is identified with the baryon number $B$. Since parity is unbroken, our states can be characterised by the parity $P$ and baryon $B$ quantum numbers i.e.\ $B^P$. The spectrum consists of mesons, baryons, glueballs and glue-fermion composite states~\cite{gluinoglue}, which can be classified by these quantum numbers and their spin. In addition to the spectral states we also measure the string tension.

On the lattice, hadron masses and in general spectral energies $m_X$ are found via the temporal asymptotic behaviour of correlators $\langle O^{\dagger}(t,{\bf x}) O(0,{\bf 0}) \rangle \buildrel t \to \infty\over\longrightarrow e^{-m_{X}t}$ where $O$ is an operator carrying the quantum numbers of the state in question. We work in the chiral representation with:
\begin{eqnarray}
	\gamma^{\mu}=\left(\begin{array}{cc}
		0 & \overline{\sigma}_{\mu}\\
		\sigma_{\mu} & 0
	\end{array}\right)\;
\ {\rm and \ charge \ conjugation \ operator} \
  {C} = i \gamma_0 \gamma_2=\left(\begin{array}{cc}
    i \sigma_2 & 0\\
    0 & - i \sigma_2
  \end{array}\right)\;, \nonumber
\end{eqnarray}
\noindent
where $\sigma^{\mu}=\left(1,{\vec \sigma}\right),\quad\overline{\sigma}^{\mu}=\left(1,-{\vec \sigma}\right)\;$ and the charge conjugation operation defined as $\psi_C = C {\overline \psi}^{\rm T}$. Possible operators $O$ bilinear in the fermion field are ${\overline \psi} {\Gamma} \psi$, ${\psi}^{\rm T} C {\Gamma} \psi$, ${\overline \psi} \Gamma C {\overline \psi}^{\rm T}$ and $\psi^{\rm T} C \Gamma {\overline \psi}^{\rm T}$. We consider $\Gamma=1, \gamma_0, {\vec \gamma}, \gamma_0 {\vec \gamma}, \gamma_{5} \gamma_{0}, \gamma_5 {\vec \gamma}, \gamma_0 \gamma_5 {\vec \gamma}, \gamma_5$. We, therefore, see that the operators correspond to meson states with $B=0$, $P=\pm$ and baryon states with $B=\pm 2$, $P=\pm$.

In addition to the mesonic and baryonic states, the theory possesses a chiral doublet that can be thought as a fermion-gluon bound state. In a supersymmetric Yang-Mills theory this state would be called gluino-glueball; here it is noted as a spin-$\frac{1}{2}$ state, since the fermions are not partners of the gluons. This state can be obtained by correlators of the lattice version of the continuum operator~\cite{gluinoglue}:
\begin{eqnarray}
O_{\textnormal{spin-}\tfrac{1}{2}}=\sum_{\mu, \nu} \sigma_{\mu \nu} {\rm tr} \left[ F^{\mu \nu} \psi_{\rm M}  \right]
\end{eqnarray}
with $\sigma_{\mu \nu}=\frac{1}{2}\left[ \gamma_{\mu},\gamma_{\nu}  \right]$. Glueball operators consist of linear combinations of ordered products of link matrices on closed paths, which have been chosen in such a way that they transform irreducibly under spin transformations, parity and charge conjugation.  

Following~\cite{su2nf2_3}, the string tension has been calculated using operators of blocked/smeared Polyakov loops and from the temporal asymptotic behaviour of expectation values of Wilson loops. The first method gives us access to the closed flux-tube spectrum, which can be fitted using an effective string theory prediction in order to extract the string tension. Wilson loops lead to the static potential. This can be fitted by a Cornell-like potential giving, thus, the string tension. 
\section{Spectrum}
\label{sec:spectrum}
\begin{figure}
	\begin{center}
	\hfill
	\includegraphics[width=0.45\textwidth]{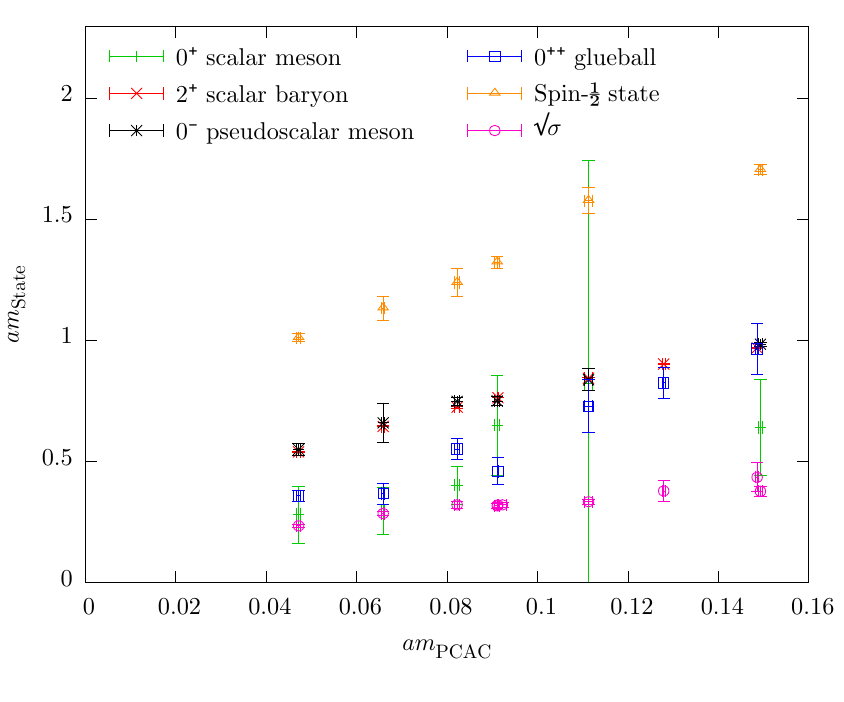}
	\hfill
        \includegraphics[width=0.45\textwidth]{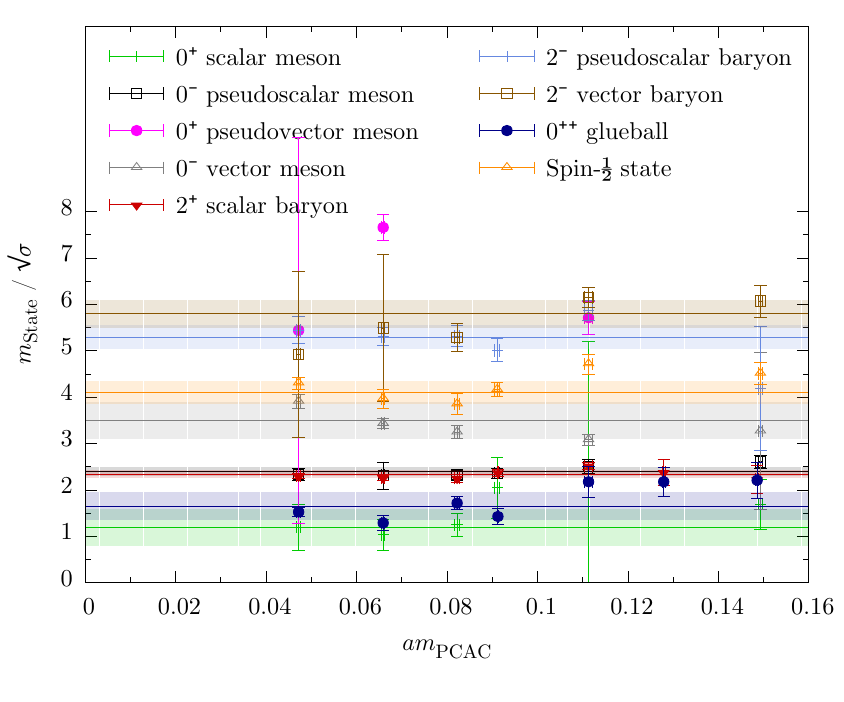}
        \hfill\null
	\end{center}
	\vspace{-24pt}
	\caption{Selected spectral states of the theory, showing meson, baryon, glueball, and spin-$\frac{1}{2}$ states, and $\sigma^{1/2}$, all tending towards 0 as $\m_{\rm PCAC} \rightarrow 0$. On the left we present the raw masses in lattice units and on the right mass ratios with the string tension. \label{fig:spectrum-unscaled}}
\end{figure}

\vspace{-0.30cm}
In Figure~\ref{fig:spectrum-unscaled} we present spectrum of several states of the theory in terms of bare quantities (left panel) and in terms of mass ratios with the string tension (right panel). In the first plot we observe that the meson, baryon, glueball and spin-$\frac{1}{2}$ masses as well as the string tension in lattice units decrease monotonically towards zero as $am_{\rm PCAC} \to 0$. Hence, there is no doubt that this theory does not follow the characteristics of the ``traditional'' confining behaviour. 

Turning now to the mass ratios, we observe that, with the exception of the pseudovector meson, over which we have a poor control, within two standard deviations all states have constant mass ratios throughout the range in which they were observed. The scalar meson $0^+$ is the lightest state in the scaling region. In addition we expect the scalar glueball $0^{++}$ and the scalar meson to mix. As a matter of fact the mass ratios of these two states appear to be degenerate within the statistical uncertainties. The mass ratios of the baryon states appear to behave in a similar manner as the mesonic states. Namely, their mass ratios are constant throughout the range of $a m_{\rm PCAC}$.

Our results, therefore, obey the hyperscaling hypothesis according to which the mass ratios of spectral quantities in the chiral scaling regime for a mass-deformed infrared conformal gauge theory should be constant. Hence, the SU(2) gauge theory with one adjoint Dirac fermion appears to possess an infrared behaviour compatible with a conformal or nearly-conformal nature of the theory. 

\vspace{-0.35cm}
\section{The chiral condensate anomalous dimension}
\label{sec:anomalous_dimension}
\begin{figure}
\begin{center}
	\includegraphics[width=0.32\textwidth]{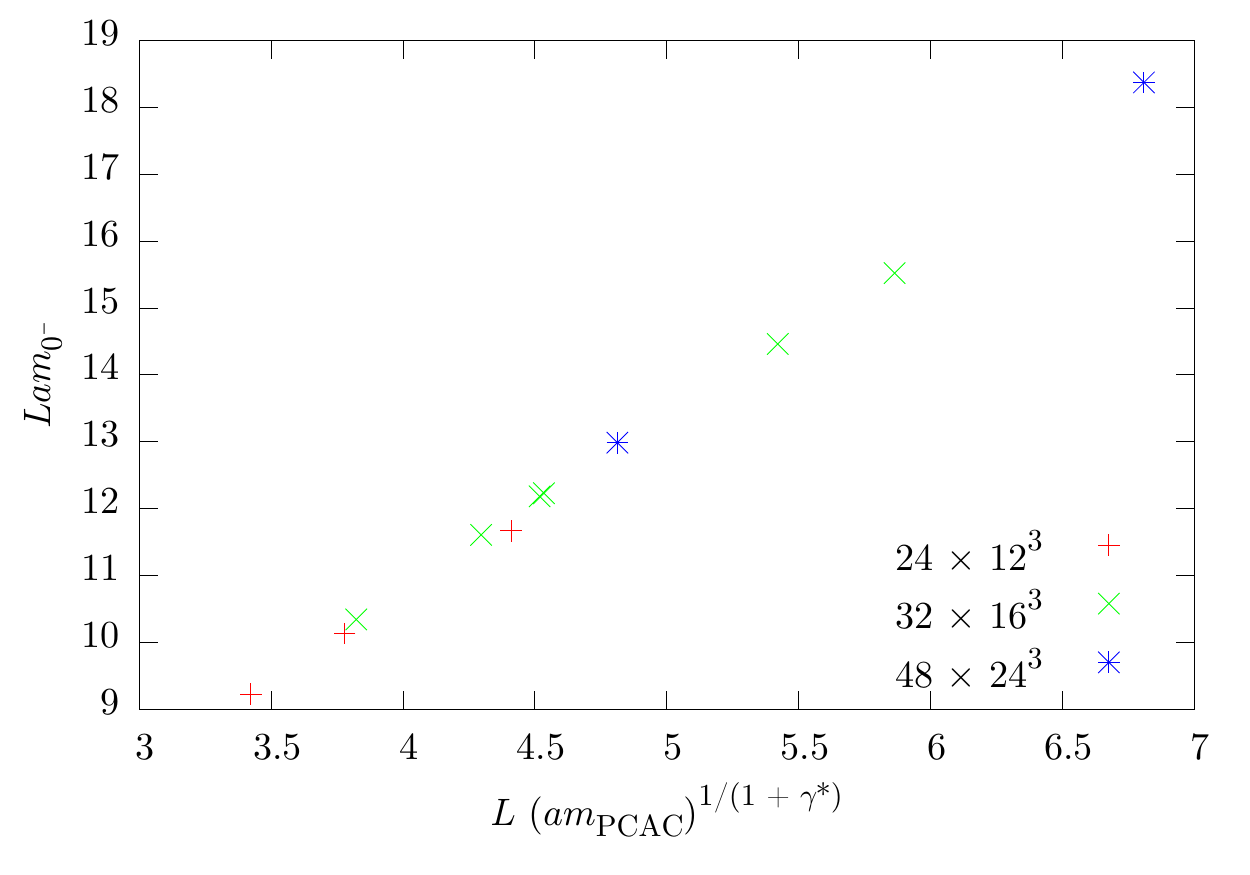} 
	\includegraphics[width=0.32\textwidth]{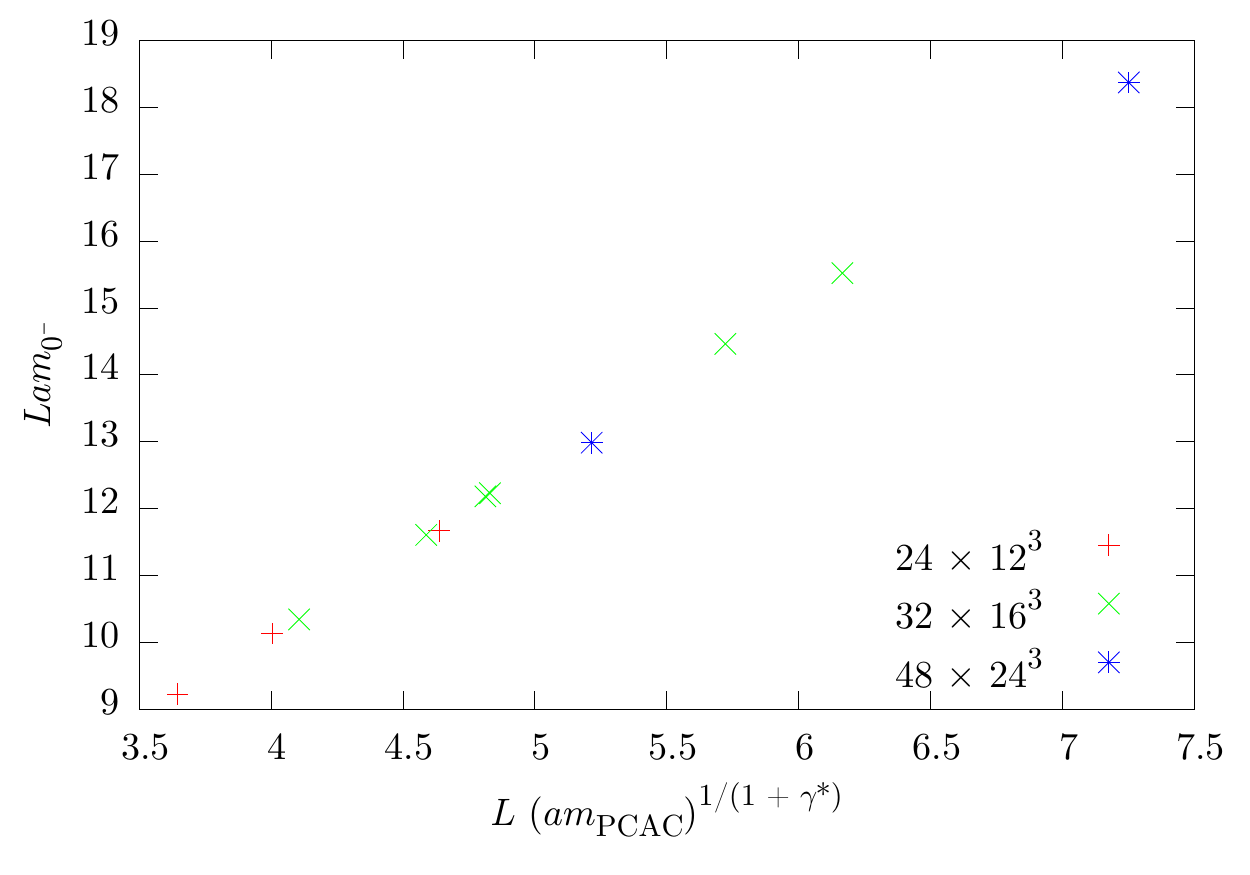} 
	\includegraphics[width=0.32\textwidth]{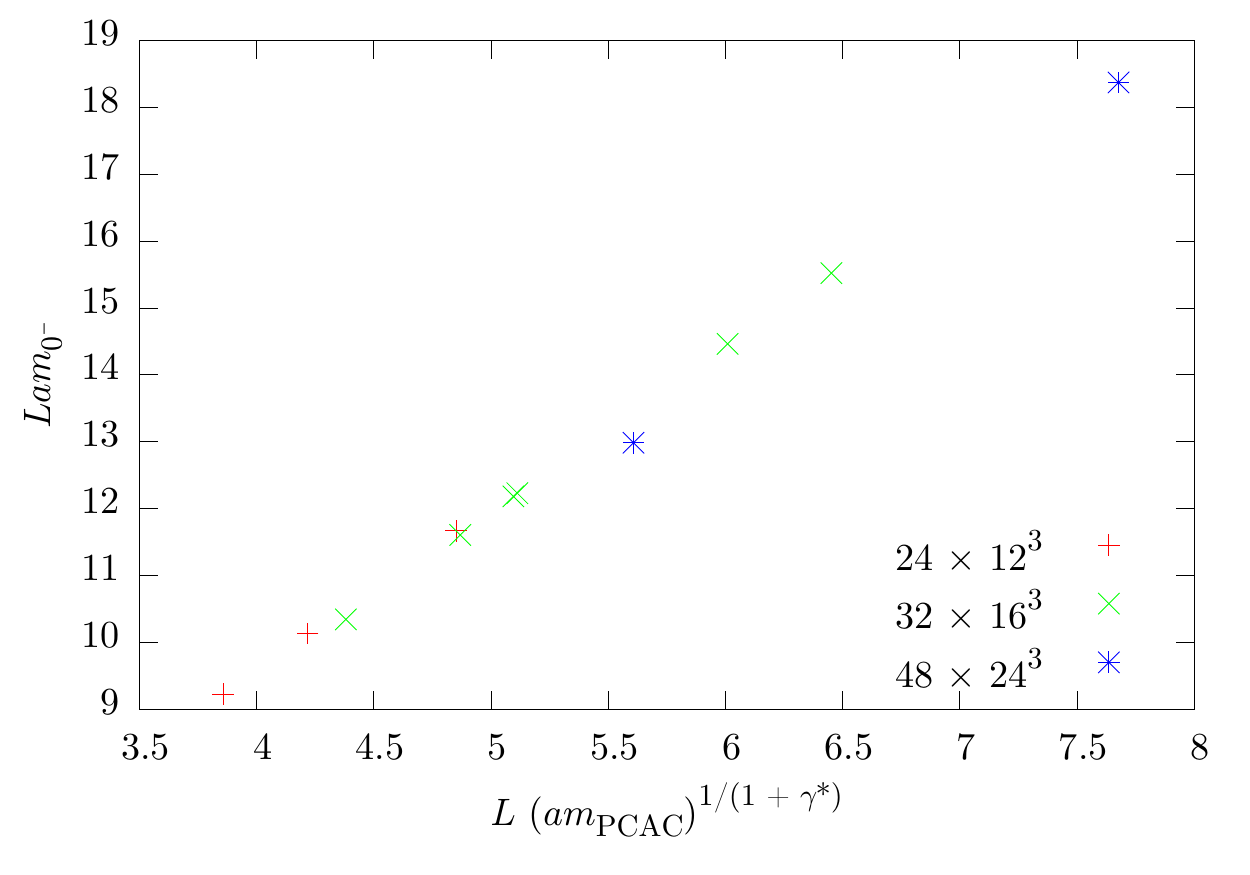}
  \end{center}
\vspace{-0.60cm}
\caption{Plots of $L m_{{0^-}}$ as a function of $L m_{\rm PCAC}^{{1}/({1+\gamma_{\ast}})} $ for the three lattice volumes $24\times12^3$, $32\times16^3$ and $48\times24^3$ and $\gamma_{\ast}=0.9, \ 1.0, \ 1.1$. The results appear to identify a universal curve for $\gamma_{\ast}= 0.9 - 1.0$.}
\label{fig:inspection}
\end{figure}

\vspace{-0.35cm}
Various techniques can be used to extract $\gamma_*$. The first method we show consists in revealing an approximate value for the anomalous dimension using finite size scaling predictions. For a conformal theory a spectral quantity $m_X$ of the system on a finite lattice of spatial extension $L$, as $L \to \infty$ and the combination $L m_{\rm PCAC}^{{1}/{(1+\gamma_{\ast}})}$ is kept constant, obeys the asymptotic formula:
\begin{eqnarray}
  L m_{X} = f \left(L m_{\rm PCAC}^{\frac{1}{1+\gamma_{\ast}}} \right),
\label{eqn:fss}
\end{eqnarray}
for some unknown function $f$. If the system is in the scaling region, the equation above can be used for determining $\gamma_{\ast}$. Hence, we plot on the same figure $ L m_{X}$ as a function of $ L m_{\rm PCAC}^{{1}/({1+\gamma_{\ast}})}$ for a fixed value of $\gamma_{\ast}$ and for several ensembles. Next, we find the suitable value of $\gamma_{\ast}$ for which the data points for different ensembles collapse on a universal curve. In Figure~\ref{fig:inspection} we provide such plots for $m_{{0^-}}$, with $\gamma_{\ast}=0.9, \ 1.0$ and $1.1$, for three different lattice volumes. These plots are taken from a sequence of plots for $\gamma_{\ast}=0.1-2.0$ with an increment of $0.1$ and demonstrate that for $\gamma_{\ast}=0.9-1.0$ our results identify a universal curve. Hence, by inspection we would expect that the anomalous dimension $\gamma_{\ast}$ would lie between 0.9 and 1.0.

\begin{figure*}
\begin{center}
  {
    \includegraphics[width=6cm]{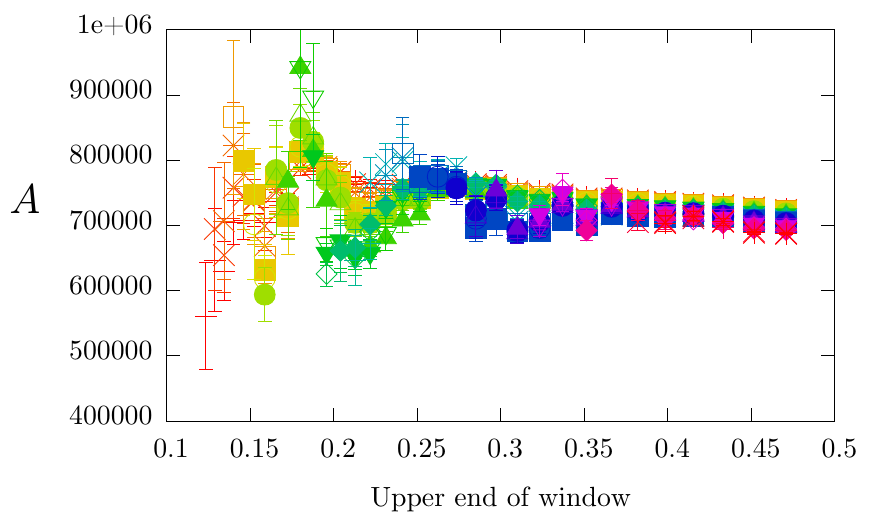}
  }
  {
    \includegraphics[width=6cm]{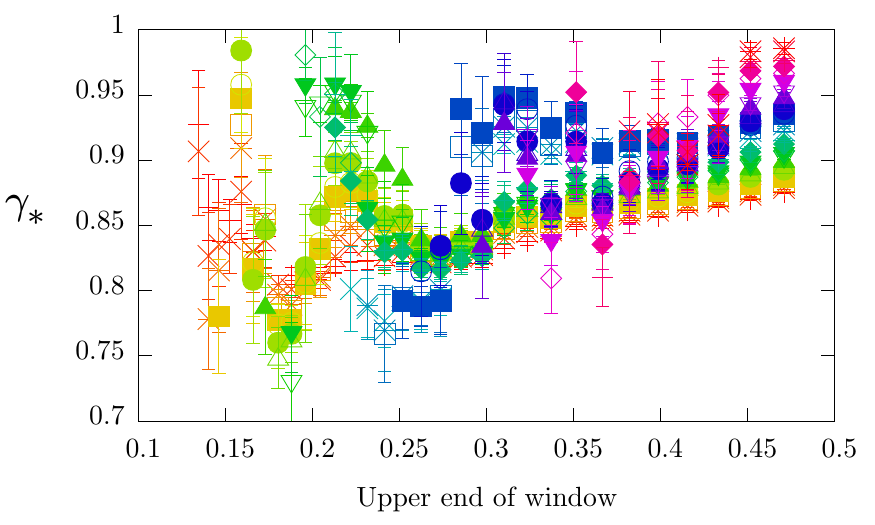}
  }

  {
    \includegraphics[width=6cm]{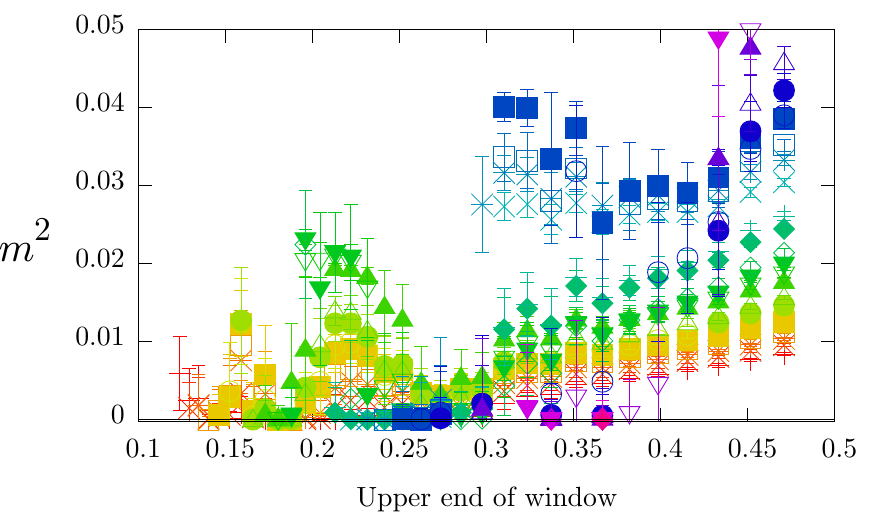}
  }
  {
    \includegraphics[width=6cm]{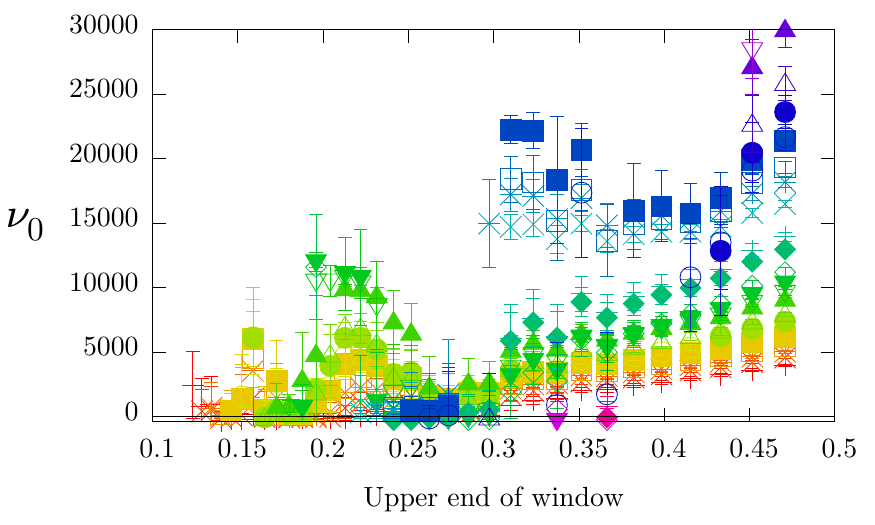}
  }
  
  \vspace{-0.25cm}
  {\scriptsize Lower end of window:}

  \vspace{0.2cm}
  {
  \includegraphics[width=\textwidth]{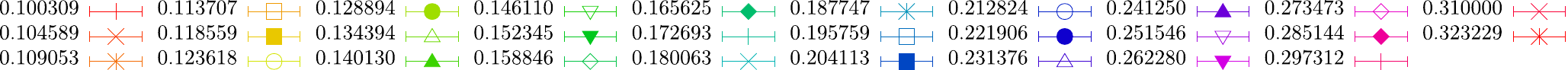}
  }
  \vspace{-1cm}
\end{center}
	\caption{Plateaux for the fitted observables for the D2 data at various lengths and positions of the fitting window. The colour represents the position of the lower end of the window, and the $x$-axis the upper end. The plateaux at the top-right of each plot were taken as the ``true'' values.\label{fig:fittings}}
\end{figure*}

Another way of obtaining $\gamma_{\ast}$, which leads to a more precise and rigorous determination, is by fitting the Dirac mode number ${\bar \nu} (\Omega)$ as a function of the Dirac eigenvalues $\Omega$~\cite{Patella}. The raw data~\cite{forthcoming} for the mode number for several lattice volumes are consistent at high $\Omega$, but diverge at low $\Omega$, moving further from a straight line as the lattice volume is reduced. It, therefore, makes more sense to focus on our results on D2. Fitting the mode number requires particular care. In fact, the scaling regime~\cite{Patella}: 
\begin{eqnarray}
   a^{-4}{\bar \nu} (\Omega) \approx a^{-4}{\bar \nu}_0(m) + A\left[ (a\Omega)^2 - (am)^2 \right]^{\frac{2}{1+\gamma_*}}
\label{eqn:modenumber}
\end{eqnarray}
is realised only in an intermediary region of $\Omega$, the extent of which is not known \emph{a priori}. In addition the fit is highly sensitive on initial conditions, creating a systematic uncertainty on the fitting parameters. For a more detailed description of the fitting method see~\cite{Patella}. In Figure~\ref{fig:fittings} we provide our results for the four fitting parameters of Equation~(\ref{eqn:modenumber}) for several ranges of the fitting window in $\Omega$ controlled by the lower and upper ends. According to~\cite{Patella} we seek for a plateau which demonstrates the stability of the fit in both ends. The best result obtained in this optimisation gives an anomalous dimension of the condensate in the range $0.9  \leq  \gamma_{\ast}  \leq  0.95$ with a best fit of $ \gamma_{\ast}= 0.92(1)$. This agrees perfectly with the result from the finite scaling scenario. Combining this result with the spectral analysis, the indication coming from our study is that the theory is conformal (or near-conformal), with an anomalous dimension of the right size for a viable EWSB model. Although, as we have discussed, the chiral symmetry breaking pattern of this model excludes it from candidate walking technicolor theories, we have preliminary evidence that large anomalous dimensions are realised near the onset of the conformal window. In order to put these conclusions on firmer grounds, a careful infinite volume extrapolation (e.g. along the lines of~\cite{deldebbio}) needs to be performed.  
   
\vspace{-0.30cm}
\section*{Acknowledgements}
\label{Acknowledgements}
\vspace{-0.30cm}
Computations for this project were performed on the DiRAC HPC facility supported by STFC, and on HPC Wales clusters supported by the ERDF through the WEFO.

\end{document}